\documentclass[prd,nofootinbib,showpacs,preprint]{revtex4-1}
\usepackage[T1]{fontenc}
\usepackage{amsmath,amssymb}
\usepackage{epsfig}
\usepackage{graphicx}
\usepackage[usenames,dvipsnames]{color}
\usepackage{slashed}
\usepackage[colorlinks,citecolor=blue]{hyperref}
\usepackage{pdfpages}
\usepackage{color}
\begin{document}
\title{Minimal Left-Right Symmetry Confronted with the 750 GeV Di-photon Excess at LHC}

\author{Arnab Dasgupta}
\email{arnab.d@iopb.res.in}
\affiliation{Institute of Physics, Sachivalaya Marg, Bhubaneshwar-751005, India}
\author{Manimala Mitra}
\email{manimala@iisermohali.ac.in}
\affiliation{Department of Physics, Indian Institute of Science Education and Research, Sector 81, SAS Nagar, Manauli 140306, India}
\author{Debasish Borah}
\email{dborah@iitg.ernet.in}
\affiliation{Department of Physics, Indian Institute of Technology Guwahati, Assam 781039, India}
\begin{abstract}
The recent results of 13 TeV  ATLAS and CMS di-photon searches show an excess at di-photon invariant mass of 750 GeV. We look for possible explanation of this within minimal left right symmetric model (MLRSM). The possible candidate is a neutral Higgs of mass 750 GeV that can decay to di-photon via charged Higgs and right handed gauge boson loop. However, the cross-section  is not consistent with the 
ATLAS and CMS results. We then discuss one possible variation of this model with universal seesaw for fermion masses that can explain this excess. 
\end{abstract}
\maketitle

\section{Introduction}
The recently reported 13 TeV center of mass energy data of the large hadron collider (LHC) experiment have pointed towards the existence of a resonance of mass about 750 GeV and width around 45 GeV decaying into two photons \cite{lhcrun2a,atlasconf,CMS:2015dxe}. The ATLAS collaboration has reported the presence of this 750 GeV resonance from their $3.2 \; \text{fb}^{-1}$ data with a statistical significance of $3.9\sigma$ ($2.3\sigma$ including look-elsewhere effects) whereas CMS collaboration has reported the same with a significance of $2.6\sigma$ from their $2.6\; \text{fb}^{-1}$ data. Apart from the mass and decay width, these two experiments have also measured the cross section $\sigma(pp\rightarrow \gamma \gamma)$ to be $10\pm3$ fb (ATLAS) and $6\pm3$ fb (CMS). The large decay width as well as the sizeable cross section have made it a challenging task to come up with beyond standard model (BSM) frameworks which can accommodate it. 

Although the reported signal could well be a statistical fluctuation, it has drawn significant attention from the particle physics community leading to a large number of interesting possible explanations including two Higgs doublet models, additional scalars coupling to vector like fermions, extra dimensions, dark matter among others as well as the implications of this signal reported in the works 
\cite{750pheno16Dec, 750GeVsinglet, 750pheno17Dec, 750pheno18Dec, 750pheno21Dec, 750pheno22Dec, 750pheno23Dec, 750pheno24Dec, LR750GeV, 750pheno25Dec, 750pheno29Dec}. In this work, we try to scrutinize one of the very popular BSM framework, known as the left right symmetric model \cite{lrsm,lrsmpot} in the light of the reported ATLAS and CMS results. Very recently, this model have been analyzed in the context of  several other excesses: such as 2.8$\sigma$  excess in the $eejj$ final state reported by CMS \cite{Khachatryan:2014dka, Gluza:2015goa}, $3.4 \sigma$ diboson excess reported by ATLAS   \cite{lrdiboson,Dobrescu:2015qna} and  \cite{ATLASdiboson} and the dijet results \cite{lrdijet} and  \cite{CMSdijet}. 

The MLRSM model has few additional Higgs states, where few of the scalars can have lower than TeV scale masses. The neutral Higgs state  which has 750 GeV mass, can decay to di-photon via charged Higgs and gauge boson loop.  We compute its production cross section  at LHC and the  branching ratio into two photons. We do a parameter scan of the model by varying the different parameters of the potential, with a fixed symmetry breaking scale. The computed cross section  $\sigma(pp \to  H^0_2 \to \gamma \gamma)$ is  way below the observed one with a few fb. We  show that the minimal version of this model with the neutral Higgs states as 750 GeV resonance can not explain the observed signal. 

We then consider the possible modification to the MLRSM in order to explain the observed signal. The easiest modification is the addition of vector like quarks which can couple directly to the 750 GeV neutral scalar. This will not only enhance the production cross section but also the partial decay width into two photons. This scenario has already been explored within several different models mentioned above.  Instead of arbitrarily adding vector like fermions into the MLRSM to explain the observed signal, we try to focus on the possibility of having these exotic fermions to serve another purposes. One possibility could be to realize these fermions within some higher fermion representations of grand unified theories like SO(10). Since TeV scale MLRSM does not give rise to gauge coupling unification at high energy scale, it will be very natural to have these exotic fermions at TeV scale which not only arise naturally within SO(10) framework but also can help to achieve gauge coupling unification. Another interesting motivation for additional vector like fermions is their role in generating masses of the known standard model fermions. This can happen for example, if the standard model Higgs can not directly couple to the left and right handed fermions of the model, but can do so only through additional heavy fermions. There is one such a version of LRSM where such vector like fermions have to be incorporated in order to generate observed fermion masses \cite{VLQlr, univSeesawLR}. This model was studied later in the context of cosmology \cite{gulr} and neutrinoless double beta decay \cite{lr0nu2beta}. The model gives rise to fermion masses through a universal seesaw framework where the standard model fermion masses arise after integrating out heavy fermions. Leaving the possibility of having new physics source within a grand unified theory framework to a future work, here we focus on the LRSM with universal seesaw for fermion masses. We consider a 750 GeV neutral singlet scalar which can couple to vector like fermions and can give rise to the observed signal. 

This paper is organized as follows. In section \ref{sec1}, we briefly discuss the MLRSM and then discuss the possibility of a 750 GeV neutral scalar in view of the LHC signal in section \ref{sec2}. In section \ref{sec3}, we briefly discuss the LRSM with universal seesaw and in section \ref{sec4} we study the possibility of explaining the LHC signal of a 750 GeV resonance decaying into two photons. We finally conclude in section \ref{sec5}.

\section{Minimal Left-Right Symmetric Model (MSLRM)}
\label{sec1}
Left-Right Symmetric Model \cite{lrsm,lrsmpot} is one of the very well motivated BSM frameworks where the gauge symmetry of the electroweak theory is extended to $SU(3)_c \times SU(2)_L \times SU(2)_R \times U(1)_{B-L}$. The right handed fermions which are singlets under the $SU(2)_L$ of SM, transform as doublets under $SU(2)_R$, making the presence of right handed neutrinos natural in this model. The Higgs doublet of the SM is replaced by a Higgs bidoublet to allow couplings between left and right handed fermions, both of which are doublets under $SU(2)_L$ and $SU(2)_R$ respectively. The enhanced gauge symmetry of the model $SU(2)_R \times U(1)_{B-L}$ is broken down to the $U(1)_Y$ of SM by the vacuum expectation value (vev) of additional Higgs scalar, transforming as triplet under $SU(2)_R$ and having non-zero $U(1)_{B-L}$ charge. This triplet  gives rise to the  Majorana masses of the  right handed neutrinos through  symmetry breaking. The heavy right handed neutrinos participate in the seesaw mechanism and generate the Majorana masses of the light neutrinos. On the other hand, the left handed Higgs triplet generates Majorana masses of  the light  neutrinos through type II.

The fermion content of the minimal LRSM is
\begin{equation}
Q_L=
\left(\begin{array}{c}
\ u_L \\
\ d_L
\end{array}\right)
\sim (3,2,1,\frac{1}{3}),\hspace*{0.8cm}
Q_R=
\left(\begin{array}{c}
\ u_R \\
\ d_R
\end{array}\right)
\sim (3^*,1,2,\frac{1}{3}),\nonumber 
\end{equation}
\begin{equation}
\ell_L =
\left(\begin{array}{c}
\ \nu_L \\
\ e_L
\end{array}\right)
\sim (1,2,1,-1), \quad
\ell_R=
\left(\begin{array}{c}
\ \nu_R \\
\ e_R
\end{array}\right)
\sim (1,1,2,-1) \nonumber
\end{equation}
Similarly, the Higgs content of the minimal LRSM is
\begin{equation}
\Phi=
\left(\begin{array}{cc}
\ \phi^0_{11} & \phi^+_{11} \\
\ \phi^-_{12} & \phi^0_{12}
\end{array}\right)
\sim (1,2,2,0)
\nonumber 
\end{equation}
\begin{equation}
\Delta_L =
\left(\begin{array}{cc}
\ \delta^+_L/\surd 2 & \delta^{++}_L \\
\ \delta^0_L & -\delta^+_L/\surd 2
\end{array}\right)
\sim (1,3,1,2), \hspace*{0.2cm}
\Delta_R =
\left(\begin{array}{cc}
\ \delta^+_R/\surd 2 & \delta^{++}_R \\
\ \delta^0_R & -\delta^+_R/\surd 2
\end{array}\right)
\sim (1,1,3,2) \nonumber
\end{equation}
where the numbers in brackets correspond to the quantum numbers with respect to the gauge group $SU(3)_c\times SU(2)_L\times SU(2)_R \times U(1)_{B-L}$. In the symmetry breaking
pattern, the neutral component of the Higgs triplet $\Delta_R$ acquires a vev to break the gauge symmetry of the LRSM into that of the SM and then to the $U(1)$ of electromagnetism by the vev of the neutral component of Higgs bidoublet $\Phi$:
$$ SU(2)_L \times SU(2)_R \times U(1)_{B-L} \quad \underrightarrow{\langle
\Delta_R \rangle} \quad SU(2)_L\times U(1)_Y \quad \underrightarrow{\langle \Phi \rangle} \quad U(1)_{em}$$
The symmetry breaking of $SU(2)_R \times U(1)_{B-L}$ into the $U(1)_Y$ of standard model can also be achieved at two stages by choosing a non-minimal scalar sector. {{We}} denote the vev of the two neutral components of the bidoublet as $k_1, k_2$ and that of triplets $\Delta_{L, R}$ as $v_{L, R}$. Considering $g_L=g_R$,  $k_2 \sim v_L \approx 0$ and $v_R \gg k_1$, the gauge boson masses after symmetry breaking can be written as 
$$ M^2_{W_L} = \frac{g^2}{4} k^2_1, \;\;\; M^2_{W_R} = \frac{g^2}{2}v^2_R $$
$$ M^2_{Z_L} =  \frac{g^2 k^2_1}{4\cos^2{\theta_w}} \left ( 1-\frac{\cos^2{2\theta_w}}{2\cos^4{\theta_w}}\frac{k^2_1}{v^2_R} \right), \;\;\; M^2_{Z_R} = \frac{g^2 v^2_R \cos^2{\theta_w}}{\cos{2\theta_w}} $$
where $\theta_w$ is the Weinberg angle. After the symmetry breaking, four neutral scalars emerge, two from the bidoublet $(H^0_0, H^0_1)$, one from right handed triplet $(H^0_2)$ and another from left handed triplet $(H^0_3)$. Similarly there are two neutral pseudoscalars, one from the bidoublet $(A^0_1)$ and another from the left handed triplet $(A^0_2)$. Among the charged scalars, there are two singly charged ones $(H^{\pm}_1, H^{\pm}_2)$ and two doubly charged ones $(H^{\pm \pm}_1, H^{\pm \pm}_2)$. 

{Under the approximations made above, the scalar masses are given by
\begin{equation}
 M^2_{H^0_0} = 2 \lambda_1 k^2_1, \, \, \,  M^2_{H^0_1} = \frac{1}{2}\alpha_3 v^2_R,  \, \, M^2_{H^0_2} = 2\rho_1 v^2_R, \, \, M^2_{H^0_3} = \frac{1}{2}(\rho_3-2\rho_1)v^2_R \end{equation}
\begin{equation}
M^2_{A^0_1} = \frac{1}{2}\alpha_3 v^2_R - 2(2\lambda_2-\lambda_3)k^2_1,  \, \, M^2_{A^0_2} = \frac{1}{2}v^2_R (\rho_3-2\rho_1),  \, \,
 M^2_{H^{\pm}_1} = \frac{1}{2} (\rho_3-2\rho_1)v^2_R +\frac{1}{4} \alpha_3 k^2_1
\end{equation}
\begin{equation}
 M^2_{H^{\pm}_2} = \frac{1}{2}\alpha_3 v^2_R + \frac{1}{4} \alpha_3 k^2_1,  \, \, 
M^2_{H^{\pm \pm}_1} = \frac{1}{2} (\rho_3-2\rho_1)v^2_R+\frac{1}{2}\alpha_3 k^2_1,  \, \, 
M^2_{H^{\pm \pm}_2} = 2\rho_2 v^2_R +\frac{1}{2} \alpha_3 k^2_1
\end{equation}
where $\lambda_i, \alpha_i, \rho_i$ are dimensionless couplings of the scalar potential of this model \cite{lrsmpot}. We take into account the following results and experimental searches that fix the dimensionless parameters. 
\begin{itemize}
\item
In the above, 
$H^0_0$ can be identified as SM like Higgs of mass 125 GeV, that fixes the  coupling $\lambda_1$.  Few  of the other couplings can be constrained after taking into account the  experimental limits on the scalar masses. 
\item
We demand that the Higgs $H^0_2$ has a mass 750 GeV that explain the di-photon bump, that fixes the dimensionless coupling $\rho_1$. 
\item
In order to avoid the flavor changing neutral currents (FCNC) processes, the neutral scalars from  bi-doublet $H^0_1, A^0_1$ have to be heavier than 10 TeV \cite{fcnc},  which puts further constraint on  $\alpha_3$ for a fixed $v_R$.  This also constrain the charged Higgs mass $H^{\pm}_2$ to be heavy. 
\item
The lower bound on the mass of doubly charged scalar $H^{\pm \pm}_1 $ from the multilepton search \cite{Aad:2014hja} fixes the coupling $\rho_3$. This automatically fixes the mass of singly charged scalar $H^{\pm}_1$ as well. 
\end{itemize}
This leaves only two free parameters $ (2\lambda_2-\lambda_3)$ and $\rho_2$ in the expressions for scalar masses to be varied arbitrarily. The other couplings in the full scalar potential are also free to be varied and are not affected by the chosen spectrum of scalar masses.}

\section{750 GeV neutral scalar in MLRSM}
\label{sec2}
Among the neutral physical scalars in MLRSM, the $H^0_0$ is the standard model like Higgs with mass $125$ GeV. The other two neutral (pseudo) scalars from the bidoublet namely $H^0_1, A^0_1$ are heavier than at least 10 TeV due to tight constraints from FCNC. However, the other three neutral scalars $(H^0_2, H^0_3, A^0_2)$ originating from the scalar triplets can lie at 750 GeV.  We consider $H^0_2 \equiv \Delta^0_R$ as a possible candidate for a 750 GeV neutral scalar decaying into two photons. 

{ As the heavy Higgs $H^0_2$ does not directly couple with gluons or quarks, hence its production will be governed by the mixing with
the SM Higgs. The production cross-section of the $750$ GeV heavy Higgs at 13 TeV LHC is \cite{physrepdjouadi}
\begin{equation}
\sigma (p p \to H^0_2) \sim  \theta^2 \times 0.85 \,\rm{pb},
\end{equation}
}
where $\theta \sim \alpha \frac{k_1}{M_{H^0_2}}$ is the mixing between SM Higgs state and the $H^0_2$, and we have considered the dimensionless parameters $\alpha_1=\alpha_2=\alpha$.

As the neutral scalar $H^0_2$  does not couple to a pair of charged fermions at the tree level, the only way it can decay into two photons is through a charged gauge or scalar boson loop. This can happen through a loop containing  $W_R$ bosons or one of the charged scalars $H^{\pm}_1, H^{\pm}_2, H^{\pm \pm}_1, H^{\pm \pm}_2$. The total production cross-section of $p p \to H^0_2 \to \gamma \gamma$ is 
\begin{equation}
\sigma(p p \to H^0_2 \to \gamma \gamma) \sim \theta^2 \times 0.85 \times \rm{Br}(H^0_2 \to \gamma \gamma) \, \rm{pb}.
\end{equation}

We first  consider the following benchmark values of the scalar masses and compute the decay widths. Following this we will provide a full parameter scan. 
\begin{equation}
 m_{H^0_0} = 125 \; \text{GeV}, \;\; m_{H^0_2} = 750 \; \text{GeV}, \;\; m_{H^{\pm}_1} = 380 \; \text{GeV}, \;\; m_{H^0_1} = m_{A^0_1} = m_{H^{\pm}_2} = 10\; \text{TeV}.
\end{equation}
\begin{equation} m_{H^{\pm \pm}_1} = 465 \; \text{GeV}, \;\; m_{H^{\pm \pm}_2} = 380 \; \text{GeV}. \end{equation}
The values are chosen in such a way to satisfy the current experimental bounds.  We show the partial decay width of $H^0_2$ to $\gamma \gamma$ and di-Higgs modes for these illustrative points in the parameter space. Few comments are in order. 

\begin{figure}[!h]
\centering
\epsfig{file=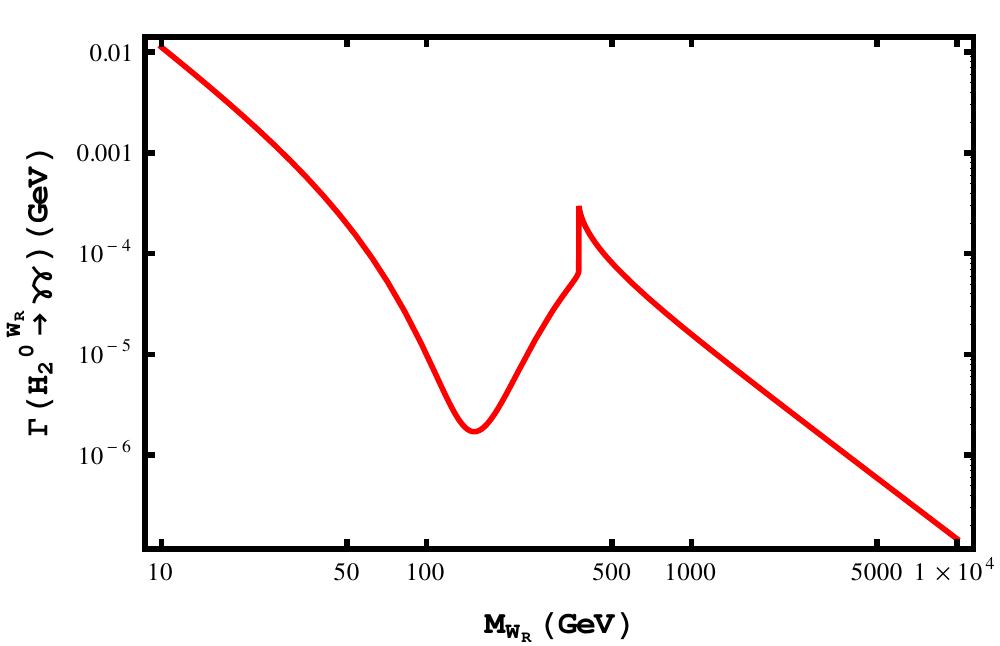,width=1.0\textwidth,clip=}
\caption{The partial decay width of $H^0_2$ to di-photon via $W_R$ loop. }
\label{fig2}
\end{figure}

\begin{figure}[!h]
\centering
\epsfig{file=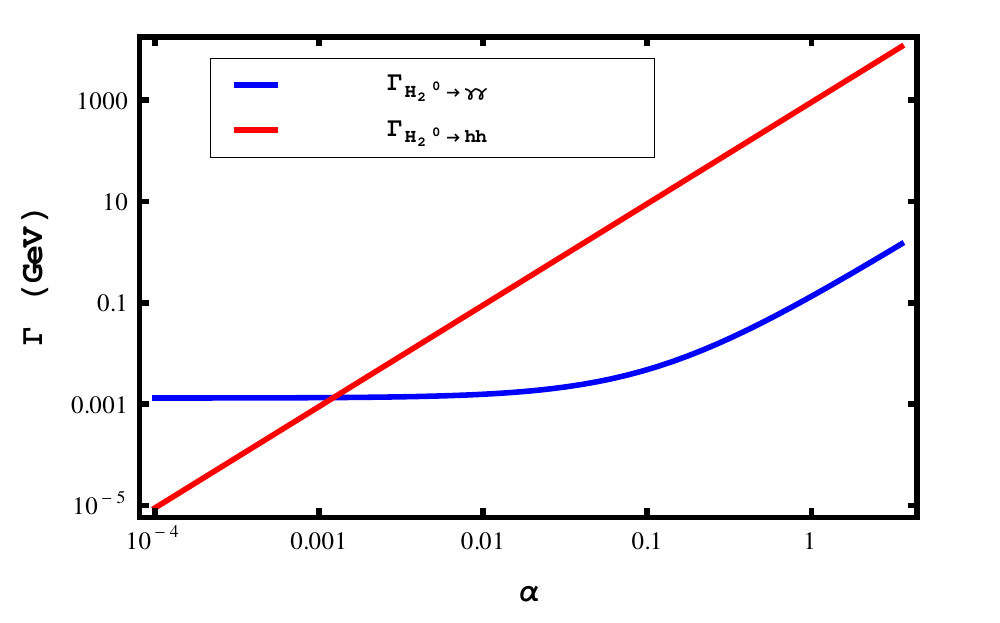,width=1.0\textwidth,clip=}
\caption{The partial decay widths  of $H^0_2$ into di-Higgs at tree level and di-photon via charged Higgs  loop.}
\label{fig1}
\end{figure}

\begin{itemize}
\item
We show the partial decay width to two photons through the $W_R$ loop in Fig.~\ref{fig2}, where the $W_R$ mass has been set to be 3 TeV-in agreement with the collider constraint \cite{Khachatryan:2014dka}. The partial decay width through gauge boson loop is extremely suppressed and can not explain
the diphoton signal. 
\item
We show the partial decay width of  $H^0_2 \rightarrow \gamma \gamma$ through charged Higgs loop in Fig.~\ref{fig1} as a function of the  parameter $\alpha$. This also decides the decay mode $H^0_2 \rightarrow H^0_0 H^0_0$. We have set  the charged Higgs masses to the lowest  possible value, which is  consistent with collider searches. The masses of the neutral scalars have been set to 10 TeV and $W_R$ mass as  3 TeV.

\item

The mixing parameter $\theta$ can be constrained from di-Higgs search \cite{dihiggscms}, which is $\mathcal{O}(0.3)$ for 750 GeV scalar resonance decaying into di-Higgs with 100$\%$ branching ratio.  

\item
From Fig.~\ref{fig1}, it is evident that for the parameter $\alpha > 0.001$, the partial decay width to di-Higgs will be larger than the di-photon. Including both the gauge boson $W_R$ and charged Higgs, this limit goes to $\alpha \sim 0.01$. 

\item
It is straightforward to see from Fig.~\ref{fig2} and Fig.~\ref{fig1}  that the partial  width of $H^0_2 \rightarrow \gamma \gamma$ through all available charged particles in loop falls below 1 GeV for small $\alpha$ and can not explain the large decay width preferred by the ATLAS and CMS data. 

\item
In the above, we have considered large mass > 1 TeV for the heavy neutrinos, which is  consistent with the collider constraint \cite{Khachatryan:2014dka}. Therefore, the decay of $H^0_2 \to N_R N_R$ is absent. However, we have checked that even with lighter heavy neutrino masses, the total decay width of $H^0_2$ is not in agreement with the di-photon results. 

\item
The partial decay width to $\gamma \gamma$ via top loop is extremely suppressed $10^{-7}$ GeV, while its decay to $t-\bar{t}$ and $W-W$  is  2.78 GeV and 11.62 GeV for mixing $\theta \sim 0.3$. 
\end{itemize}

{{ The total  cross section $\sigma (pp\rightarrow H^0_2 \rightarrow \gamma \gamma) = \sigma(pp\rightarrow H^0_2) \text{BR}(H^0_2 \rightarrow \gamma \gamma)$ for the above mentioned parameter values is $ < $ 1 fb for smaller branching ratio, and clearly can not take into account the required cross-section $\sim $ 10 fb to fit the ATLAS and CMS data. }} Considering the other two neutral scalars $H^0_3$ and $A^0_2$ as potential 750 GeV candidate will not significantly improve the situation. This requires beyond MLRSM physics to explain the recently observed di-photon excess at 750 GeV by ATLAS and CMS. 

\begin{figure}[!h]
\centering
\epsfig{file=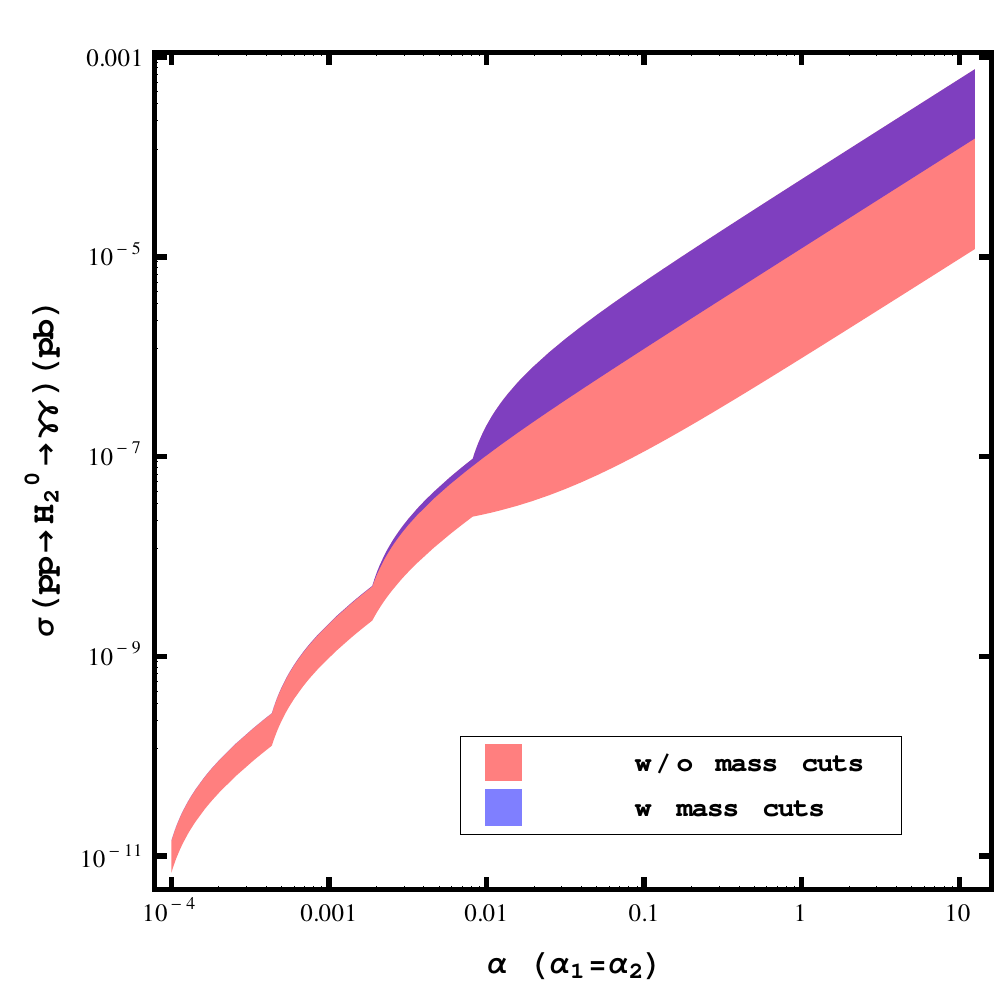,width=0.70\textwidth,clip=}
\caption{The total cross-section of $p p \to H^0_2 \to \gamma \gamma$ vs the mixing parameter $\alpha$ at 13 TeV LHC. The blue region is after imposing the mass cut given in Eq.~\ref{masscut}.}
\label{figx}
\end{figure}

\begin{figure}[!h]
\centering
\epsfig{file=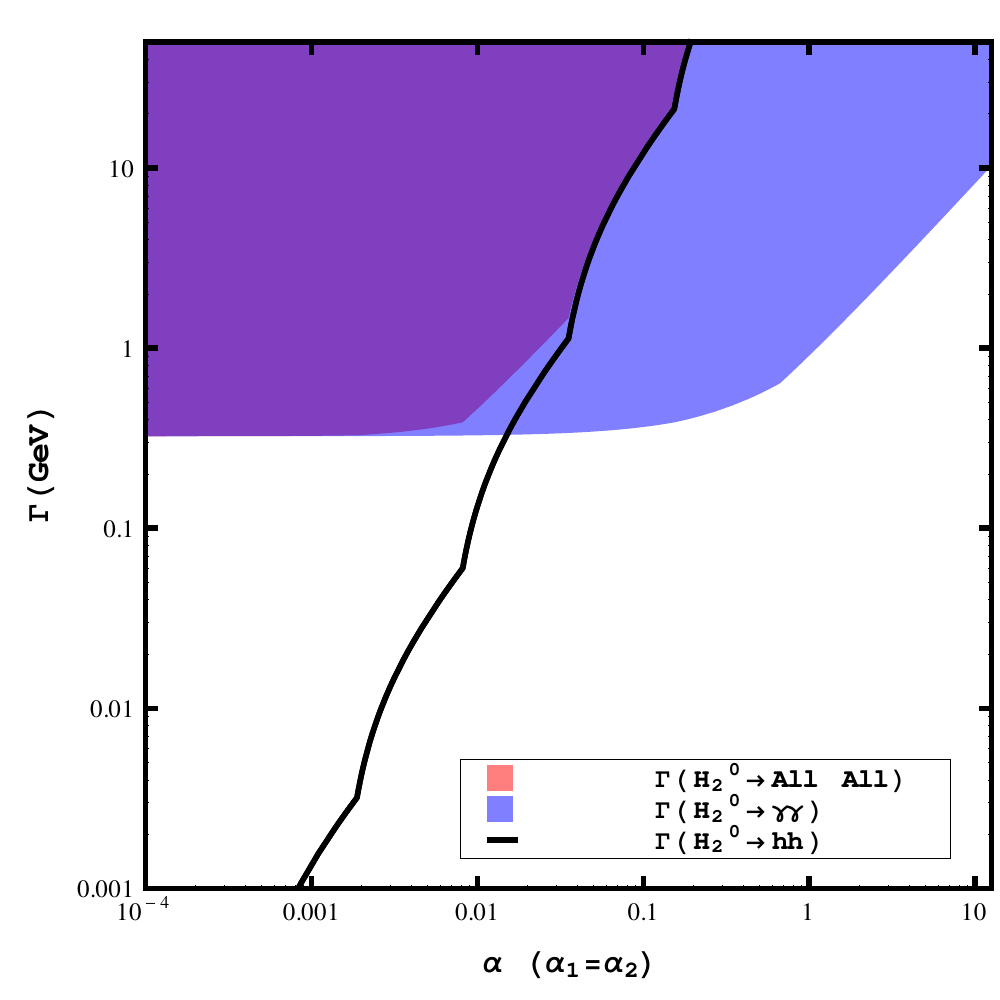,width=0.70\textwidth,clip=}
\caption{The partial as well as total decay width of the heavy Higgs $H^0_2$ vs the mixing parameter $\alpha$, after imposing the mass cut given in Eq.~\ref{masscut}. }
\label{figdw}
\end{figure}

{ {Following the above discussion with a particular set of parameters, we now provide a full parameter scan where we vary the parameters $ \alpha_1=\alpha_2 , \alpha_3,  \rho_3,  \rho_2, 2\lambda_2-\lambda_3$ in the following ranges.}}
\begin{equation}
\alpha_1  , \alpha_3,  \rho_3,  \rho_2, 2\lambda_2-\lambda_3 \equiv 10^{-4} - 4 \pi.
\end{equation}

In Fig.~\ref{figx} and Fig.~\ref{figdw}, we show the total cross-section and decay width for these ranges of parameters with $\alpha_1=\alpha_2=\alpha$. The blue  region corresponds to the following mass cut on the scalars:
\begin{equation}
 m_{H^{\pm}_{1}}, m_{H^{\pm \pm}_{2}} \ge 380 \; \text{GeV}, \;\; m_{H^0_1} = m_{A^0_1} = m_{H^{\pm}_2} > 10\; \text{TeV}, \, \, m_{H^{\pm \pm}_1} \ge 465 \; \text{GeV}.
\label{masscut}
\end{equation}
Note that, with the cut on the  decay width of $H^0_2 \le 50 $ GeV, the total cross-section is extremely small  and can not explain the 
di-photon result. In addition to this, although for large value of $\alpha$, the cross-section increases, however the branching ratio to di-Higgs also increases. Hence to avoid any constraint from di-Higgs channel, $\alpha$ should be small.

\section{LRSM with Universal Seesaw (LRSM-US)}
\label{sec3}
The fermion content of the LRSM with universal seesaw is the extension of the MLRSM fermion content by the following vector like fermions 
$$ U_{L} (3, 1, 1, \frac{4}{3}), \;\; U_{R} (3^*, 1, 1, \frac{4}{3})\;\; D_{L} (3, 1, 1, -\frac{2}{3}), \;\;D_{R} (3^*, 1, 1, -\frac{2}{3}) $$
$$ E_{L,R} (1,1,1, -2), \;\; N_{L,R} (1,1,1,0) $$
each of which comes in three different copies corresponding to the three fermion generations of the MLRSM or the standard model. The presence of these extra fermions is necessary due to the fact that the usual scalar sector of the MSLRM is replaced by the following scalar fields
$$ H_L (1, 2, 1, -1), \;\;H_R (1,1,2,-1), \;\; \sigma (1,1,1,0) $$
Due to the absence of the usual bidoublet, the left and right handed fermion doublets of the MSLRM can not directly couple to each other. However, they can couple to the scalar fields $H_{L,R}$ via the additional vector like fermions. 
\begin{align}
\mathcal{L} & \supset Y_U (\bar{Q_L} H^{\dagger}_L U_L+\bar{Q_R} H^{\dagger}_R U_R) + Y_D (\bar{Q_L} H_L D_L+\bar{Q_R} H_R D_R) +M_U \bar{U_L} U_R+ M_D \bar{D_L} D_R\nonumber \\
& Y_E (\bar{\ell_L} H_L E_L+\bar{\ell_R} H_R E_R) +M_E \bar{E_L} E_R+ \text{h.c.}
\end{align}
where we have ignored the terms corresponding to neutrino masses. For details of the origin of neutrino masses, one may refer to the discussions in \cite{gulr}. After integrating out the heavy fermions, the charged fermions of the standard model develop Yukawa couplings to the scalar doublet $H_L$ as follows
$$ y_u = Y_U \frac{v_R}{M_U} Y^T_U, \;\;y_d = Y_D \frac{v_R}{M_D} Y^T_D, \;\;y_e = Y_E \frac{v_R}{M_E} Y^T_E $$
where $v_R$ is the vev of the neutral component of $H_R$. The apparent seesaw then can explain the observed mass hierarchies among the three generations of fermions.  The non-zero vev of the neutral component of $H_R$ also breaks the $SU(2)_R\times U(1)_{B-L}$ symmetry of the model into $U(1)_Y$ of the standard model. The left handed Higgs doublet can acquire a non-zero vev $v_L$ at a lower energy to induce electroweak symmetry breaking. However, the left-right symmetry of the theory forces one to have the same vev for both $H_L$ and $H_R$ that is, $v_L=v_R$ which is unacceptable from phenomenological point of view. To decouple these two symmetry breaking scales, the extra singlet scalar $\sigma$ is introduced into the model. This field is odd under the discrete left-right symmetry and hence couple to the two scalar doublets with a opposite sign. After this singlet acquires a non-zero vev at high scale, this generates a difference between the effective mass squared of $H_L$ and $H_R$ which ultimately decouples the symmetry breaking scales. 
\section{750 GeV neutral scalar in LRSM-US}
\footnote{While preparing this manuscript, we found a similar model proposed by \cite{LR750GeV} with additional field content.}
\label{sec4}
The LRSM with universal seesaw has three neutral scalars: one from $H_L$, one from $H_R$ and one from the singlet $\sigma$. Now the neutral scalar part of $H_L$ is the standard model like Higgs with mass 125 GeV. On the other hand the neutral scalars from $H_R$ is supposed to be heavy in order to avoid dangerous flavor changing neutral currents. The scalar $\sigma$ is naturally heavy as it is responsible for discrete left-right symmetry breaking at a scale above $v_R$. To allow the possibility of a neutral 750 GeV scalar, we add another singlet $\zeta$ into the model which can couple to the vector like quarks and leptons as $Y_{f\zeta} \zeta \bar{f_L}f_R$ where $f$ is the vector like fermion. This essentially boils down to the singlet scalar resonance coupled to additional vector like fermions as an explanation of 750 GeV di-photon excess put forward by \cite{750GeVsinglet}. 

\begin{figure}[!h]
\centering
\epsfig{file=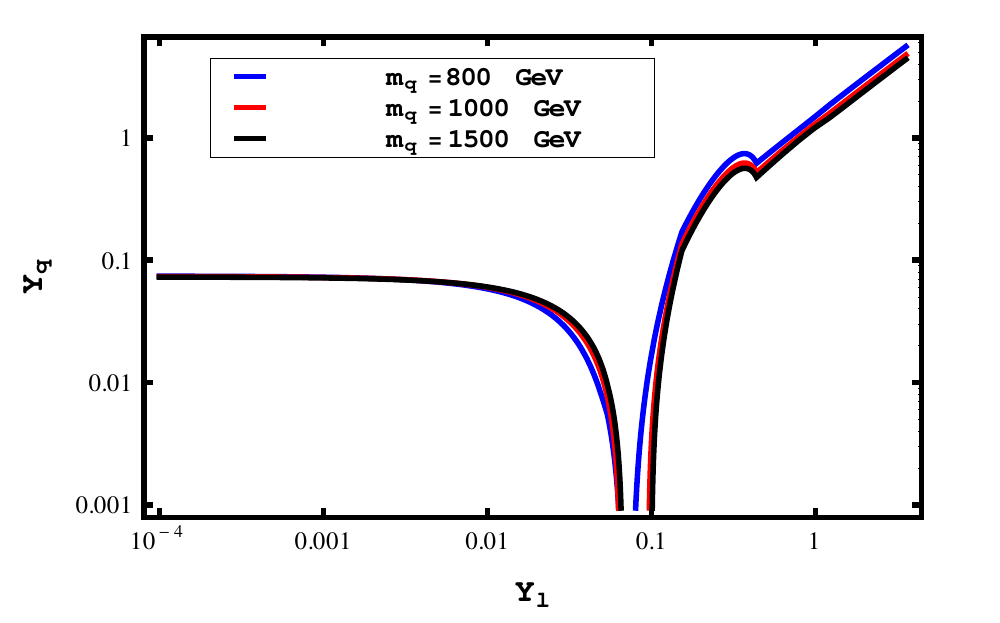,width=1.0\textwidth,clip=}
\caption{Vector like quark and vector like lepton couplings to 750 GeV scalar $\zeta$ which gives $\sigma(pp\rightarrow \zeta \rightarrow \gamma \gamma) = 10$ fb.}
\label{figyukawa}
\end{figure}

The singlet scalar can be produced dominantly in pp collisions by two different ways: (a) through mixing with the standard model Higgs and (b) through gluon gluon fusion via new vector like quarks. The singlet scalar can decay into two photons through the vector like fermions in a loop. Since the mixing with the standard model Higgs is constrained, we assume the corresponding production channel to be negligible. We then consider the production of the singlet scalar $\zeta$ in proton proton collisions dominantly through gluon gluon fusion with the vector like quarks in loop. This singlet scalar can decay either into two photons or two gluons or one photon, one Z boson at one loop level whereas the tree level decay into a pair of standard model Higgs can be neglected assuming small mixing. Using the loop level production cross section and decay width expressions given in \cite{physrepdjouadi}, we calculate for what values of vector like fermion masses $m_f$ and their Yukawa couplings  $Y_f$, the desired cross section $\sigma(pp\rightarrow \zeta \rightarrow \gamma \gamma)$ can be obtained. For simplicity we consider all the quark and lepton masses and their Yukawa couplings degenerate. Since the masses of vector like leptons are less constrained than that of vector like quarks, we consider vector like lepton masses to be half of vector like quark masses.  It should be noted that vector like quark masses are restricted to be $m_q \geq 750-920$ GeV depending on the particular channel of decay \cite{VLQconstraint} whereas this bound gets relaxed to $m_q \geq 400$ GeV \cite{VLQconstraint2} for long lived vector like quarks. Further constraints on vector like quarks can be found in \cite{vlqhandbook}. The constraints on vector like leptons are much weaker $m_l \geq 114-176$ GeV and allows the possibility of the 750 GeV scalar to decay into them at tree level \cite{VLLconstraint}. We however, do not allow tree level decay of $\zeta$ into vector like leptons which will reduce the branching ratio $\text{BR} (\zeta \rightarrow \gamma \gamma)$. Considering $m_l = m_q/2$, we then constrain the corresponding Yukawa couplings $Y_q, Y_l$ from the requirement of producing the observed signal. The restricted Yukawa couplings for some benchmark values of quark masses are shown in figure \ref{figyukawa}.

Further constraints on the model comes in terms of the Yukawa couplings involved in the seesaw relations for fermion masses discussed in the previous section. For standard model fermion mass $m_f$, the Yukawa couplings are constrained as 
\begin{equation}
\frac{y^2 v_R}{M} = \frac{m_f}{v_L}
\end{equation}
where $M$ is the heavy vector like fermion mass, $v_R$ is the $SU(2)_R$ breaking scale and $v_L=246$ GeV is the electroweak symmetry breaking scale. If $m_f$ is top quark mass and $v_R \approx 6$ TeV, then for $M=1500$ GeV, the corresponding Yukawa couplings are constrained to be $y \approx 0.42$. However, fitting with all the fermion masses will require non-degenerate heavy vector like fermion masses. Another constraints comes from the mixing of these heavy fermions with the standard model fermions. For the case of vector like quarks, such mixings with standard model quarks are constrained to be small $\theta \leq 0.1$ from precision measurements of electroweak parameters \cite{vlqhandbook}. Parametrising the heavy-light quark mixing from the seesaw relations as $\sin^2\theta \approx \frac{y^2 v_R v_L}{M^2}$ and using the constraints $\theta \leq 0.1, \frac{y^2 v_R v_L}{M} = m_f$, we get the constraints on the heavy quark masses as $M \geq 100 m_f$. This is possible to achieve for $M=1500$ GeV in case of bottom and lighter quark masses. But for top quark seesaw, the corresponding heavy vector like quark has to be much heavier than 1500 GeV considered in this simple analysis. 

It should be noted that there are 6 vector like quarks and 3 vector like charged leptons in the model.  However, the singlet scalar can not decay into them at tree level. These exotic fermions only appear at one loop to allow the scalar to decay into $gg, \gamma \gamma$ or $\gamma Z$. However, such loop level decay is not enough to generate the total decay width observed by the LHC. Similar observations were also made by \cite{750GeVsinglet}. If the LHC confirms the measured decay width in future, this will invite further modification to the left-right model considered in this work.
\section{ Conclusion}
\label{sec5}
We have studied the minimal left right symmetric extension of the standard model in view of the latest LHC observations of a 750 GeV neutral resonance decaying into two photons with a cross section of around 10 fb. Since the extra neutral scalars (in addition to the 125 GeV Higgs boson) from the bidoublet of MLRSM are very heavy to be in agreement with flavor constraints, we consider the neutral scalar $H^0_2$ from one of the triplet scalars of the model namely, $\Delta_R$. The discussion will be similar for the neutral component of $\Delta_L$. Since the triplet does not couple to quarks, we consider the production of this scalar only through its mixing with the standard model Higgs. We then consider the possible decay of $H^0_2$ and calculate the total as well as partial decay widths. After incorporating the LHC constraints on neutral as well as charged scalar masses, we find that the total cross section $\sigma(pp\rightarrow H^0_0 \rightarrow \gamma \gamma)$ remain below the observed 10 fb signal, after putting constrain on  the decay width  of $H^0_2 \le 50$ GeV. The cross section is maximal, close to 1 fb only for very high values of the dimensionless parameter $\alpha = \alpha_1 = \alpha_2$ of the scalar potential. This parameter also decides the size of the $H^0_2$ mixing with the standard model like Higgs $H^0_0$ and hence constrained to be $\alpha \leq 0.3 M_{H^0_2}/k_1$, to avoid collider constraint. Thus, the di-photon cross section  will be much smaller than 1 fb after taking the constraint on $H^0_0-H^0_2$ mixing into account. It is observed from Fig.~\ref{figx}, that after taking the experimental lower bounds on neutral and charged scalar masses into account, the total cross section gets shifted to higher side. This is due to the fact that, experimental lower bounds on scalar masses also restricts relevant dimensionless parameters to high values for fixed $v_R$. As the same parameters also appear in the decay widths, they increases the total cross section. We also observe that the neutral scalar $H^0_2$ can have a sizeable total decay width $\approx 50$ GeV as observed by the LHC for allowed parameter space though it can not give rise to the 10 fb di-photon signal simultaneously.  

We then briefly mention another possible left right model with universal seesaw for fermion masses. Due to the existence of additional vector like fermions, the production of a neutral scalar and its decay into two photons can he enhanced at the same time. By taking some benchmark values of additional fermion masses, we show how their couplings to a neutral 750 GeV scalar get restricted from the requirement of producing a 10 fb di-photon signal. The neutral scalar in such a scenario however, fails to give rise to the large decay width observed by experiment. Thus, if the di-photon cross section as well as the decay width are both confirmed by future LHC data, then further improvement of the left right symmetric models discussed in this work will be required. We leave such an investigation to future work.

\begin{acknowledgments}
The work of M.M was supported by the  DST-INSPIRE Faculty  grant. The authors  would like to thank Ketan M. Patel, IISER Mohali for very useful discussions.
\end{acknowledgments}

\end{document}